\documentstyle[11pt,IAUS212,twoside,epsf]{article}

\def\edcomment#1{\iffalse\marginpar{\raggedright\sl#1\/}\else\relax\fi}
\reversemarginpar

\begin{document}
\vspace*{1cm}
\title{Spectroscopic analysis of newborn massive stars in SMC N81}
 \author{F. Martins \& D. Schaerer}
\affil{Laboratoire d'Astrophysique, Observatoire Midi-Pyr\'en\'ees, 14 Av. Edouard Belin, F-31400 Toulouse, France}
\author{M. Heydari-Malayeri}
\affil{LERMA, Observatoire de Paris, 61 Av. de l'Observatoire, F-75014 Paris, France}

\begin{abstract}
We present the first results of a spectroscopic study of young massive stars in the SMC High Excitation Blob N81. 
These stars have M$_{v}$ values which are $\sim$ 2 mag smaller than those of normal dwarf 
stars. Their UV STIS spectra reveal features typical of O stars, 
but surprisingly weak wind components. The preliminary modelling of these spectra with the code CMFGEN 
(Hillier \& Miller 1998) indicates a mass 
loss rate of the order $10^{-9}$ M$_{\odot}$/yr. If confirmed, such a weak wind may indicate either a breakdown of the 
wind-momentum luminosity relation at low luminosity, or a steeper slope of this relation at low metallicity. 
\end{abstract}

\section{Introduction}
N81 belongs to the class of the 
High Excitation Blobs (HEB) first introduced by Heydari-Malayeri \& Testor (1982) (see Heydari-Malayeri et al., 
these proceedings). It displays cavities, 
shocks, ionisation ridges, turbulent structures typical of massive star forming environments. Several stars 
just emerging from their parental cloud are grouped within the 2'' constituting the core region.  Thus, 
N81 represent a unique opportunity to study both the earliest phases of the evolution of massive stars and the 
metallicity dependence of their wind properties.

\section{Results}
A qualitative analysis of the STIS spectra (Fig. 1) reveals that the N81 stars are O dwarfs (presence of
NV($\lambda1240$), OV($\lambda1371$), CIV($\lambda1550$) and HeII($\lambda1640$); absence of strong SiIV($\lambda1400$)) 
probably lying near the ZAMS as indicated by their low luminosity and the weakness of their wind features 
(see Heydari-Malayeri et al 2002) : they likely belong to the Vz class (Walborn \& Parker 1992).

The preliminary quantitative analysis of one of the two brightest stars of N81 using the non-LTE spherically 
expanding line blanketed code CMFGEN (Hillier \& Miller 1998) gives : $T_{eff} = 40000$ K, $\log g = 4.1$, $\log 
(L/L_{\odot}) = 5.15$, $\log \dot{M} = -9.0$, $v_\infty=2000$ km/s. Fig. 1 shows the wind momentum - luminosity relation 
(WLR) for O stars : the N81 star has a wind $\sim$ 2 orders of magnitude weaker than normal dwarfs. This may 
indicate either a breakdown of the WLR at low luminosity, or a steeper slope at low metallicity.

\begin{figure}
\plottwo{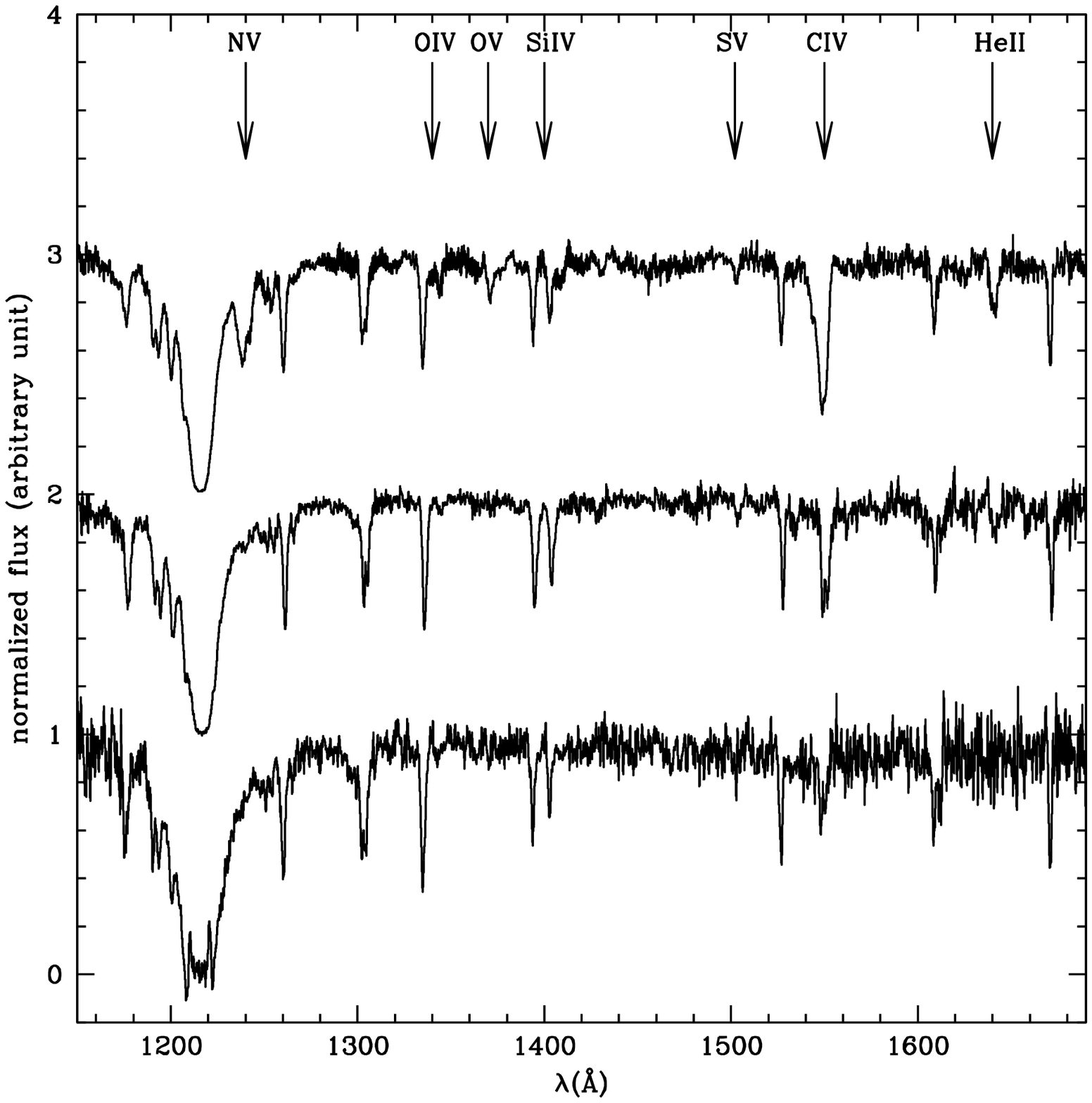}{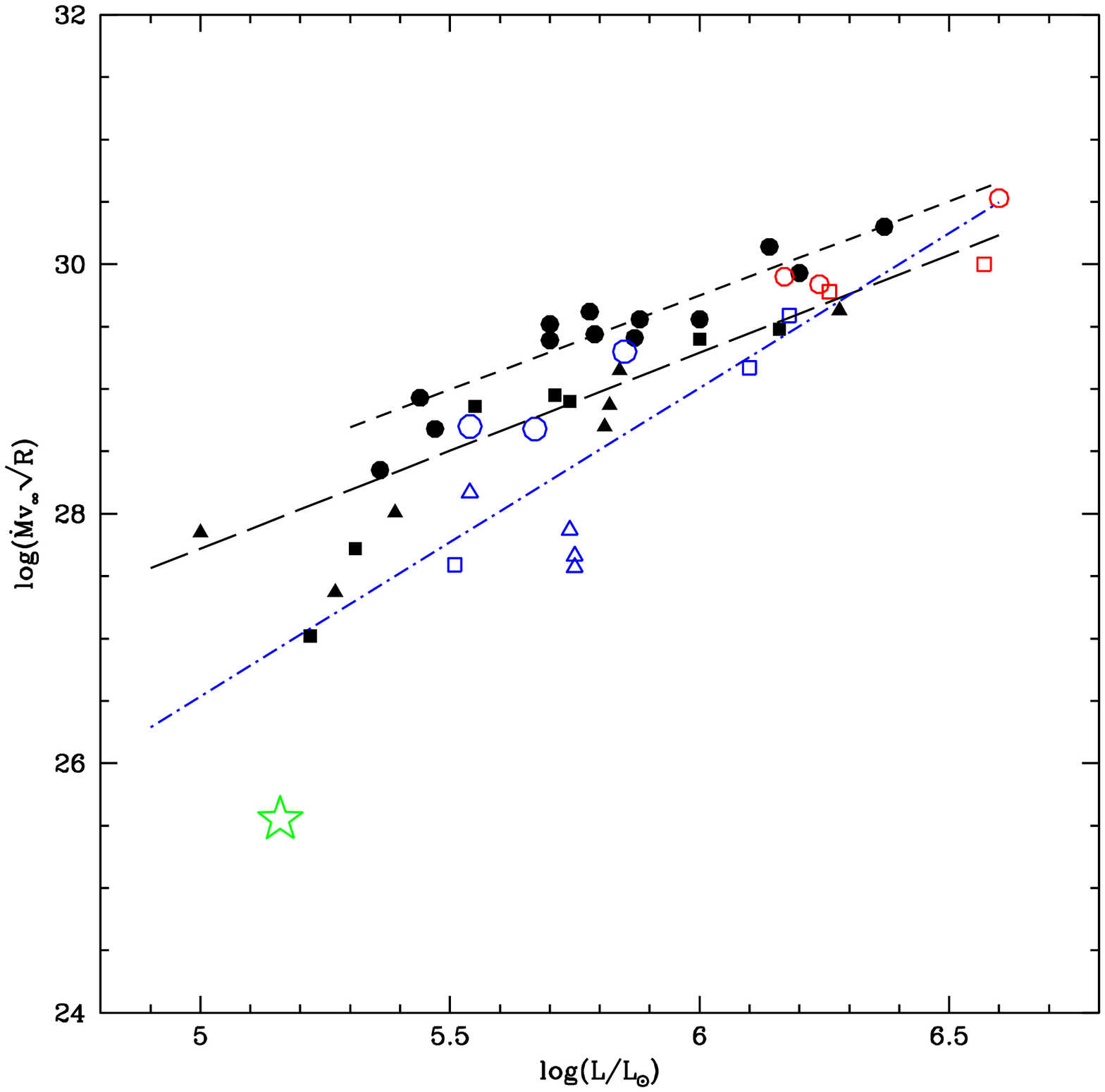}
\caption{Left panel : STIS spectra of 3 N81 stars. 
Right panel : Wind Momentum - Luminosity relation for O stars (filled (open) symbols : Galaxy (LMC \& SMC)) 
from Puls et al (1996), Herrero et al (2000),  
Hillier et al. (2002), Crowther et al. (2002). Triangles (squares, circles) are for luminosity class V (III,I).
Regression curves for galactic dwarfs (long dashed), galactic supergiants (short dashed) (Kudritzki \& Puls 2000) 
and SMC stars (Lamers \& Cassinelli 1996, dot dashed) are shown. The star symbol is our N81 star.}
\end{figure}

\section{Conclusion}
Work is in progress to improve our determination of the parameters of these N81 O stars. If confirmed, such weak winds 
for O stars are puzzling because they are not consistent with the prediction of radiation driven winds theory (e.g. Vink 
et al. 2000, Hoffmann et al, Puls, these proceedings).

\pagebreak

\end{document}